\documentclass[conference]{IEEEtran}
\IEEEoverridecommandlockouts
\usepackage{cite}
\usepackage{amsmath, amssymb, amsfonts}
\usepackage[caption=false,font=scriptsize,labelfont=sf,textfont=sf]{subfig}
\usepackage{algorithmic}
\usepackage{graphicx}
\usepackage{textcomp}
\usepackage{xcolor}
\usepackage[bottom = 3.0cm,top = 1.9cm,left=1.6cm,right=1.6cm]{geometry}
\def\BibTeX{{\rm B\kern-.05em{\sc i\kern-.025em b}\kern-.08em
 T\kern-.1667em\lower.7ex\hbox{E}\kern-.125emX}}
\begin{document}

\title{\LARGE Stacked Intelligent Metasurfaces for Multiuser Beamforming\\in the Wave Domain}

\author{Jiancheng An, Marco Di Renzo, M\'erouane Debbah, and Chau Yuen
\thanks{This research is supported by the Ministry of Education, Singapore, under its MOE Tier 2 (Award number MOE-T2EP50220-0019). Any opinions, findings and conclusions or recommendations expressed in this material are those of the author(s) and do not reflect the views of the Ministry of Education, Singapore. The work of M. Di Renzo was supported in part by the European Commission through the H2020 ARIADNE project under grant agreement number 871464 and through the H2020 RISE-6G project under grant agreement number 101017011.}
\thanks{J. An, and C. Yuen are with the Engineering Product Development Pillar, Singapore University of Technology and Design, Singapore 487372 (e-mail: jiancheng\_an@sutd.edu.sg; yuenchau@sutd.edu.sg). M. Di Renzo is with Universit\'e Paris-Saclay, CNRS, CentraleSup\'elec, Laboratoire des Signaux et Syst\`emes, 91192 Gif-sur-Yvette, France (e-mail: marco.di-renzo@universite-paris-saclay.fr). M. Debbah is with the Technology Innovation Institute, 9639 Masdar City, Abu Dhabi, United Arab Emirates, and also with CentraleSup\'elec, Universit\'e Paris-Saclay, 91192 Gif-sur-Yvette, France (email: merouane.debbah@tii.ae).}}
\maketitle

\begin{abstract}
Reconfigurable intelligent surface has recently emerged as a promising technology for shaping the wireless environment by leveraging massive low-cost reconfigurable elements. Prior works mainly focus on a single-layer metasurface that lacks the capability of suppressing multiuser interference. By contrast, we propose a stacked intelligent metasurface (SIM)-enabled transceiver design for multiuser multiple-input single-output downlink communications. Specifically, the SIM is endowed with a multilayer structure and is deployed at the base station to perform transmit beamforming directly in the electromagnetic wave domain. As a result, an SIM-enabled transceiver overcomes the need for digital beamforming and operates with low-resolution digital-to-analog converters and a moderate number of radio-frequency chains, which significantly reduces the hardware cost and energy consumption, while substantially decreasing the precoding delay benefiting from the processing performed in the wave domain. To leverage the benefits of SIM-enabled transceivers, we formulate an optimization problem for maximizing the sum rate of all the users by jointly designing the transmit power allocated to them and the analog beamforming in the wave domain. Numerical results based on a customized alternating optimization algorithm corroborate the effectiveness of the proposed SIM-enabled analog beamforming design as compared with various benchmark schemes. Most notably, the proposed analog beamforming scheme is capable of substantially decreasing the precoding delay compared to its digital counterpart.
\end{abstract}


\section{Introduction}
Reconfigurable intelligent surface (RIS) has recently emerged as a promising technology for effectively improving the spectrum and energy efficiencies of wireless networks \cite{TCOM_20202_An_Low, JSAC_2020_Renzo_Smart, TWC_2019_Huang_Reconfigurable, Proc_2022_Alexandropoulos_Pervasive}. In general, an RIS is made of a programmable metasurface consisting of a large number of low-cost passive elements integrated with low-power electronics, each of which can independently impose a phase shift on the incident waves \cite{TC_2021_Wu_Intelligent, WCL_2022_An_Scalable}. By adjusting the phase shifts of all the elements with the aid of a smart controller, an RIS is capable of manipulating the reflected/transmitted waves to shape the wireless channels, thus establishing favorable propagation environments and substantially improving the quality-of-service (QoS) of wireless networks \cite{WC_2020_Huang_Holographic, TCOM_2020_Wu_Beamforming, TWC_2020_Guo_Weighted, TWC_2022_Papazafeiropoulos_Intelligent, TGCN_2022_An_Joint, JSTSP_2022_Wei_Multi, JWCN_2019_Renzo_Smart}.

Specifically, by deploying an RIS, \emph{Huang et al.} \cite{TWC_2019_Huang_Reconfigurable} achieved 300\% energy efficiency improvement in multiuser multiple-input single-output (MISO) downlink communication systems, as compared with conventional amplify-and-forward relays. Moreover, \emph{Wu et al.} \cite{TC_2021_Wu_Intelligent} demonstrated that a $6$ dB power gain is attained upon doubling the number of reflecting elements. In order to reduce the excessive pilot signaling overhead required for probing RIS-aided channels with a large number of programmable elements, \emph{An et al.} \cite{WC_2022_An_Codebook, TCOM_20202_An_Low} proposed a novel codebook-based protocol, striking a beneficial trade-off between the QoS, the pilot overhead, and the computational complexity. However, existing works on RIS-aided wireless systems mostly rely on a single-layer metasurface structure, which severely limits the degrees-of-freedom of the attainable beam patterns \cite{TCOM_2020_Wu_Beamforming, TGCN_2022_An_Joint} and the capability of suppressing the multiuser interference \cite{TWC_2020_Guo_Weighted}.

Recently, a wave-based signal processing paradigm relying on a multilayer metasurface structure has gained prominent attention \cite{Science_2018_Lin_All, NE_2022_Liu_A}. Specifically, \emph{Lin et al.} \cite{Science_2018_Lin_All} conceived a diffractive deep neural network (D$^2$NN) architecture, by stacking an array of printed optical lenses, which is capable of performing parallel calculations at the speed of light \cite{Science_2018_Lin_All}. Subsequently, \emph{Liu et al.} \cite{NE_2022_Liu_A} designed a programmable D$^2$NN architecture based on multi-layer metasurfaces, where each meta-atom acts as a reprogrammable artificial neuron. The authors demonstrated that the programmable D$^2$NN architecture is able to execute various signal processing tasks, e.g., image classification, by flexibly manipulating the electromagnetic (EM) waves propagating through it \cite{NE_2022_Liu_A}.

In wireless communications, however, the design of transceivers based on multilayer metasurfaces remains largely unexplored. Motivated by these considerations, we develop a new stacked intelligent metasurface (SIM) enabled wireless transceiver design. As depicted in Fig. \ref{fig_1}, by stacking multiple programmable metasurfaces in the close proximity of the transmitter, an SIM has a structure that is similar to an artificial neural network and, therefore, offers enhanced signal processing capabilities compared to its single-layer counterpart. Benefiting from the resulting processing capabilities of SIMs, we consider a scenario in which an SIM is deployed at the base station (BS) to perform multiuser downlink beamforming directly in the EM wave domain. In sharp contrast to the recent multi-layer RIS that requires complex baseband signal processing capabilities and power-hungry radio-frequency (RF) chains at the transceiver \cite{TCOM_2022_Liu_Compact}, the proposed SIM-enabled transceiver completely removes the digital beamforming and only requires low-resolution digital-to-analog converters (DAC) at the BS.

Specifically, we propose a wave-based (analog) beamforming scheme for application to an SIM-assisted multiuser MISO downlink system. First of all, we formulate a sum rate maximization problem by jointly optimizing the transmit power and the wave-based beamforming. Since the transmit power coefficients and the large number of SIM phase shifts are coupled in a non-convex objective function, we propose an effective alternating optimization (AO) algorithm to solve the formulated joint optimization problem. Numerical results demonstrate the effectiveness of the novel wave-based beamforming paradigm and the performance improvement of the proposed AO algorithm against various benchmark schemes.
\section{System Model}\label{sec2}
\begin{figure}[!t]
\centering
\includegraphics[width=7cm]{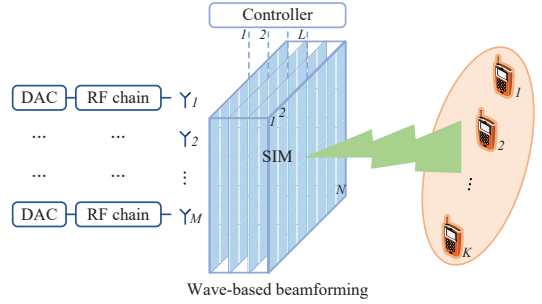}
\caption{An SIM-assisted multiuser MISO downlink system.}\vspace{-0.6cm}
\label{fig_1}
\end{figure}

As shown in Fig. \ref{fig_1}, we consider an SIM-aided multiuser MISO wireless system, where an SIM is deployed to assist the downlink communication from a BS equipped with $M$ antennas to $K$ single-antenna users. Specifically, the SIM is composed of $L$ metasurface layers, each of which consists of $N$ meta-atoms. Moreover, the SIM is attached to a smart controller, which is capable of imposing an independent and adjustable phase shift onto the EM waves transmitted through each meta-atom \cite{TCOM_2020_Wu_Beamforming}. By properly adjusting the phase shifts in each metasurface, the SIM implements the downlink beamforming directly in the EM wave domain \cite{NE_2022_Liu_A}.

Let $\mathcal{L} = \left \{ 1, 2, \cdots , L \right \}$, $\mathcal{N} = \left \{ 1, 2, \cdots , N \right \}$, and $\mathcal{K} = \left \{ 1, 2, \cdots , K \right \}$ denote the set of metasurfaces, meta-atoms in each metasurface layer, and users, respectively. Moreover, let $e^{j\theta _{n}^{l}},\, \forall n \in \mathcal{N},\, \forall l \in \mathcal{L}$ with $\theta _{n}^{l} \in \left [ 0,2\pi \right )$ denote the phase shift imposed by the $n$-th meta-atom in the $l$-th metasurface layer. Thus, the diagonal phase shift matrix $\mathbf{\Phi}^{l}$ of the $l$-th metasurface layer is $\mathbf{\Phi}^{l} = \text{diag}\left ( e^{j\theta _{1}^{l}}, e^{j\theta _{2}^{l}}, \cdots , e^{j\theta _{N}^{l}} \right ) \in \mathbb{C}^{N\times N},\, \forall l\in \mathcal{L}$. Furthermore, let $\mathbf{W}^{l}\in \mathbb{C}^{N\times N},\, \forall l \neq 1,\, l \in \mathcal{L}$ be the transmission matrix from the $\left ( l-1 \right )$-th to the $l$-th metasurface layer and let $\mathbf{w}_{m}^{1}\in \mathbb{C}^{N\times 1}$ denote the transmission vector from the $m$-th transmit antenna to the first metasurface layer of the SIM. According to the Rayleigh-Sommerfeld diffraction theory \cite{Science_2018_Lin_All}, the $\left ( n,{n}' \right )$-th entry $w_{n,{n}'}^{l}$ of $\mathbf{W}^{l}$ is given by
\begin{small}\begin{align}\label{eq3}
 w_{n,{n}'}^{l} =\frac{d_{x}d_{y}\cos\chi _{n,{n}'}^{l} }{d_{n,{n}'}^{l}}\left ( \frac{1}{2\pi d_{n,{n}'}^{l}}-j\frac{1}{\lambda } \right )e^{j2\pi d_{n,{n}'}^{l}/\lambda },
\end{align}\end{small}for $\forall l \neq 1,\, l \in \mathcal{L}$, where $\lambda$ is the wavelength, $d_{n,{n}'}^{l}$ denotes the transmission distance, $\chi _{n,{n}'}^{l}$ represents the angle between the propagation direction and the normal direction of the $\left ( l-1 \right )$-th metasurface layer, and $d_{x} \times d_{y}$ is the size of each meta-atom. Similarly, the $n$-th entry $w_{n,m}^{1}$ of $\mathbf{w}_{m}^{1}$ is obtained from \eqref{eq3}. Therefore, the SIM-enabled wave-based beamforming matrix $\mathbf{G}$ is written as
\begin{small}\begin{align}\label{eq4}
 \mathbf{G}=\mathbf{\Phi }^{L}\mathbf{W}^{L}\mathbf{\Phi }^{L-1}\cdots \mathbf{\Phi }^{2}\mathbf{W}^{2}\mathbf{\Phi }^{1}\in \mathbb{C}^{N\times N}.
\end{align}\end{small}It is worth noting that the inter-layer transmission coefficients may deviate from \eqref{eq3} due to practical hardware imperfections, which, however, can be readily calibrated before the deployment of the SIM \cite{NE_2022_Liu_A}.

As a sensible case study, we consider a quasi-static flat-fading channel model. Specifically, $\mathbf{h}_{k}^{H} \in \mathbb{C}^{1\times N},\, \forall k \in \mathcal{K}$ denotes the baseband equivalent channel from the last metasurface layer of the SIM to the $k$-th user, which is modeled as a correlated Rayleigh fading channel, i.e., $\mathbf{h}_{k}\sim \mathcal{CN}\left ( \mathbf{0}, \beta _{k}\mathbf{R} \right ),\, \forall k\in \mathcal{K},$ where $\beta _{k}$ is the distance-dependent path loss of the $k$-th SIM-user link, and $\mathbf{R}\in \mathbb{C}^{N\times N}$ is the covariance matrix that characterizes the spatial correlation between different meta-atoms. By assuming an isotropic scattering environment with uniformly distributed multipath components, the $\left ( n,{n}' \right )$-th entry of $\mathbf{R}$ is $\mathbf{R}_{n,{n}'}=\text{sinc} \left ( 2d_{n,{n}'}/\lambda \right )$ \cite{WCL_2021_Bjornson_Rayleigh}, where $\text{sinc}\left ( x \right ) = \sin\left ( \pi x \right )/\left ( \pi x \right )$ is the normalized sinc function, and $d_{n,{n}'}$ denotes the spacing between the meta-atoms.

In sharp contrast with conventional digital precoding schemes that assign each symbol to an individual beamforming vector, we consider a wave-based beamforming scheme with the aid of the SIM. Accordingly, each data stream is directly transmitted from the BS antennas. In general, the number of BS antennas $M$ can be greater than the number of users $K$, and an antenna selection process needs to be carried out in advance \cite{CM_2004_Sanayei_Antenna}. For simplicity, we assume $M = K$. In the considered case study, each antenna performs the modulation and detection of a single data stream. Therefore, the use of low-resolution ADC/DAC may be a suitable choice, while resulting in acceptable performance losses.

The information symbol intended to the $k$-th user is denoted by $s_k,\, \forall k\in \mathcal{K}$, which is assumed to be an independent and identically distributed (i.i.d.) random variable with zero mean and unit variance. Let $p_{k} \geq 0$ denote the power allocated to the $k$-th user. Then, the total transmit power constraint at the BS reads as $\sum_{k = 1}^{K}p_{k} \leq P_{T},$ where $P_{T}$ is the transmit power budget at the BS. Furthermore, by superimposing all the signals that propagate through the SIM, the composite signal $y_{k}$ received at the $k$-th user is expressed as
\begin{small}\begin{align}
 y_{k} = \mathbf{h}_{k}^{H}\mathbf{G}\sum\limits_{{k}' = 1}^{K}\mathbf{w}_{{k}'}^{1}\sqrt{p_{{k}'}}s_{{k}'} + n_{k}, \ \forall k \in \mathcal{K},
\end{align}\end{small}where $n_{k}\sim \mathcal{CN}\left ( 0,\sigma _{k}^{2} \right )$ denotes the i.i.d. additive white Gaussian noise with $\sigma _{k}^{2}$ being the receiver noise power at the $k$-th user.

Therefore, the signal-to-interference-plus-noise-ratio (SINR) at the $k$-th user is given by
\begin{small}\begin{align}\label{eq9}
 \gamma _{k} = \frac{\left | \mathbf{h}_{k}^{H}\mathbf{G}\mathbf{w}_{k}^{1} \right |^{2}p _{k}}{\sum\nolimits_{{k}'\neq k}^{K}\left | \mathbf{h}_{k}^{H}\mathbf{G}\mathbf{w}_{{k}'}^{1} \right |^{2}p _{{k}'} + \sigma _{k}^{2}}, \ \forall k \in \mathcal{K}.
\end{align}\end{small}

In this paper, we are interested in optimizing the phase shifts in $\mathbf{G}$ to suppress the multiuser interference, which we refer to as the \emph{wave-based beamforming}. Compared with conventional digital beamforming and RIS-aided hybrid active and passive beamforming schemes, the considered wave-based beamforming scheme entails the optimization of a greater number of phase shifts associated with each metasurface layer, which is a more challenging problem to solve as elaborated in the next section.

\section{Joint Power Allocation and Wave-Based Beamforming: Problem Formulation and Solution}\label{sec4}
\subsection{Problem Formulation}
We aim to maximize the sum rate of all the users, by jointly optimizing the transmit power at the BS and the wave-based beamforming at the SIM. To characterize the ultimate performance limits, we assume that the CSI of all the channels is perfectly known by the BS\footnote{In the considered SIM-aided system model, channel estimation can be performed similarly to conventional MIMO systems with a reduced number of RF chains, e.g., \cite{TCOM_2016_Lee_Channel}.}. By defining $\mathbf{p} \triangleq \left [ p_{1},p_{2},\cdots ,p_{K} \right ]^{T}$, ${\boldsymbol{\theta }}^{l} \triangleq \left [ \theta _{1}^{l}, \theta _{2}^{l}, \cdots , \theta _{N}^{l} \right ]^{T}$, and $\boldsymbol{\vartheta } \triangleq \left \{ \boldsymbol{\theta} ^{1},\boldsymbol{\theta} ^{2},\cdots ,\boldsymbol{\theta} ^{L} \right \}$, the joint power allocation and wave-based beamforming optimization problem is formulated as
\begin{small}\begin{subequations}
\begin{alignat}{3}
\left ( P1 \right ): \quad & \max_{\mathbf{p},\, \boldsymbol{\vartheta }} \quad && R=\sum \nolimits_{k = 1}^{K}\log_{2}\left ( 1+\gamma _{k} \right ) \label{eq11a}\\
{} \quad & \text{s.t.} \quad && \sum\nolimits_{k=1}^{K}p_{k}\leq P_{T}, \label{eq11b}\\
{} \quad & {} \quad && p_{k}\geq 0,\ \forall k \in \mathcal{K}, \label{eq11c}\\
{} \quad & {} \quad && \theta _{n}^{l} \in \left [ 0,2\pi \right ),\ \forall n \in \mathcal{N},\ \forall l \in \mathcal{L}. \label{eq11d}
\end{alignat}
\end{subequations}\end{small}Since the optimization variables $\mathbf{p}$ and $\boldsymbol{\vartheta }$ are coupled in the non-convex objective function in \eqref{eq11a}, it is not simple to obtain the optimal solution of problem $\left ( P1 \right )$. To tackle this issue, we propose an efficient AO algorithm.

\subsection{AO Algorithm}
The proposed AO algorithm decomposes the joint optimization problem $\left ( P1 \right )$ into a pair of subproblems: the power allocation problem among $K$ interfering channels at the BS, and the phase shift optimization problem at the SIM. At any iteration of each subproblem, we update one of the optimization variables, i.e., $\mathbf{p}$ or $\boldsymbol{\vartheta }$, while keeping the other one fixed to the value that it takes at the previous iteration.

\subsubsection{Algorithm 1 for Optimizing the Power Allocation $\mathbf{p}$ Given $\boldsymbol{\vartheta }$}\label{sec41}
Assuming that the SIM phase shifts $\boldsymbol{\vartheta } = \left \{ \boldsymbol{\theta} ^{1},\boldsymbol{\theta} ^{2},\cdots ,\boldsymbol{\theta} ^{L} \right \}$ and the wave-based beamforming matrix $\mathbf{G}$ in \eqref{eq4} are given, the problem $\left ( P1 \right )$ reduces to
\begin{small}\begin{subequations}
\begin{alignat}{3}
\left ( P2 \right ): \quad & \max_{\mathbf{p}} \quad && R\\
{} \quad & \text{s.t.} \quad && \eqref{eq11b},\ \eqref{eq11c}
\end{alignat}
\end{subequations}\end{small}The problem $\left ( P2 \right )$ is a conventional power allocation problem over $K$ interference channels with a given total power constraint. Therefore, it can be efficiently solved by applying the iterative water-filling algorithm \cite{TIT_2005_Jindal_Sum}.

Specifically, the power allocated to the $k$-th user at each iteration is \cite{TCOM_20202_An_Low}
\begin{small}\begin{align}\label{eq13}
 p_{k}=\left ( p_{o}-\frac{\sum\nolimits_{{k}'\neq k}^{K}\left | \mathbf{h}_{k}^{H}\mathbf{G}\mathbf{w}_{{k}'}^{1} \right |^{2}p _{{k}'} + \sigma _{k}^{2}}{\left | \mathbf{h}_{k}^{H}\mathbf{G}\mathbf{w}_{k}^{1} \right |^{2}} \right )^{+},
\end{align}\end{small}for $\forall k\in \mathcal{K}$, where $\left ( x \right )^{+} \triangleq \max\left \{ 0,x \right \}$, and $p_{o}$ is the water-filling level that is determined by using the bisection method while fulfilling the constraint $\sum\nolimits_{k=1}^{K}p_{k}=P_{T}$ at each iteration. 

Since the power allocated to all the users is simultaneously updated at each iteration, the iterative water-filling algorithm specified by \eqref{eq13} may not be stable when $K > 2$ \cite{TIT_2005_Jindal_Sum}. To circumvent this issue, we add a damping term. Specifically, at each iteration, the power allocation $\mathbf{p}$ is a weighted sum of the power at the previous iteration and the power in \eqref{eq13}. This ensures that the sum rate $R$ approaches its maximum value after several iterations and computations of \eqref{eq13}.
\subsubsection{Algorithm 2 for Optimizing the SIM Phase Shifts\label{sec42} $\boldsymbol{\vartheta }$ Given $\mathbf{p}$}
Given the power allocation $\mathbf{p}$, the phase shift optimization subproblem is formulated as
\begin{small}\begin{subequations}
\begin{alignat}{3}
\left ( P3 \right ): \quad & \max_{\boldsymbol{\vartheta }} \quad && R\\
{} \quad & \text{s.t.} \quad && \eqref{eq11d} \label{eq15b}
\end{alignat}
\end{subequations}\end{small}The problem $\left ( P3 \right )$ is still challenging to solve due to the non-convex objective function. To tackle it, we propose a computationally efficient iterative gradient ascent algorithm.

First, the phase shifts $\theta _{n}^{l},\, \forall n\in \mathcal{N},\, \forall l \in \mathcal{L}$ are initialized by assuming, e.g., a uniform random distribution. Then, at each iteration, the partial derivative of $R$ with respect to $\theta _{n}^{l}$ is calculated as follows
\begin{small}\begin{align}\label{eq15}
 \frac{\partial R}{\partial \theta _{n}^{l}}&=2\log_{2}e\sum_{k=1}^{K}\delta _{k} \left ( p_{k}\eta _{k,k}-\gamma _{k}\sum\limits_{{k}'\neq k}^{K}p_{{k}'}\eta _{k,{k}'} \right ),
\end{align}\end{small}
where $\delta _{k}$ and $\eta _{k,{k}'}$ are given by
\begin{small}\begin{align}
 \delta _{k}&=\frac{1}{\sum\nolimits_{{k}' = 1}^{K}\left | \mathbf{h}_{k}^{H}\mathbf{G}\mathbf{w}_{{k}'}^{1} \right |^{2}p _{{k}'} + \sigma _{k}^{2}},\\
 \eta _{k,{k}'}&=\text{Im} \left [ e^{-j\theta _{n}^{l}}\left ( \mathbf{w}_{{k}'}^{1} \right )^{H}\mathbf{u}_{n}^{l}\left ( \mathbf{v}_{n}^{l} \right )^{H}\mathbf{h}_{k}\mathbf{h}_{k}^{H}\mathbf{G}\mathbf{w}_{{k}'}^{1} \right ].\label{eq11}
\end{align}\end{small}In \eqref{eq11}, $\left ( \mathbf{u}_{n}^{l} \right )^{H}$ and $\mathbf{v}_{n}^{l}$ denote the $n$-th row of the matrix $\mathbf{U}^{l}\in \mathbb{C}^{N\times N}$ and the $n$-th column of the matrix $\mathbf{V}^{l}\in \mathbb{C}^{N\times N}$, which are defined by
\begin{small}\begin{align}
\mathbf{U}^{l}&\triangleq \begin{cases}
 \mathbf{W}^{l}\mathbf{\Phi }^{l-1}\cdots \mathbf{\Phi }^{2}\mathbf{W}^{2}\mathbf{\Phi }^{1}, & \text{if}\ l \neq 1,\\
 \mathbf{I}_{N}, & \text{if}\ l=1,
 \end{cases}\\
\mathbf{V}^{l}&\triangleq\begin{cases}
 \mathbf{\Phi }^{L}\mathbf{W}^{L}\mathbf{\Phi }^{L-1}\cdots \mathbf{\Phi }^{l+1}\mathbf{W}^{l+1}, & \text{if}\ l\neq L,\\
 \mathbf{I}_{N}, & \text{if}\ l=L.
 \end{cases}
\end{align}\end{small}

After calculating all the partial derivatives of $R$ with respect to $\theta _{n}^{l}$ according to \eqref{eq15}, all the phase shifts $\left \{ \theta _{n}^{l} \right \}$ of the SIM are simultaneously updated as follows
\begin{small}\begin{align}\label{eq14}
 \theta _{n}^{l}\leftarrow \theta _{n}^{l}+\mu \frac{\partial R}{\partial \theta _{n}^{l}}, \ \forall n \in \mathcal{N}, \ \forall l \in \mathcal{L},
\end{align}\end{small}where $\mu > 0$ is the \emph{Armijo} step size, which is obtained by leveraging the backtracking line search at each iteration \cite{TWC_2022_Papazafeiropoulos_Intelligent}.

The computation of \eqref{eq14} is repeated until the fractional increase of the sum rate is less than a preset threshold, i.e., until convergence. The convergence of the gradient ascent algorithm to a local maximum is guaranteed because 1) the sum rate $R$ is upper bounded due to the constraint on the transmit power and the unit values of the amplitudes of the transmission coefficients of the SIM; and 2) the sum rate $R$ is non-decreasing provided that a suitable \emph{Armijo} step size $\mu$ is utilized at each iteration.

By alternatively executing Algorithms 1 and 2 several times, the solution to problem $\left ( P1 \right )$ is obtained.

\section{Numerical Results}\label{sec6}
\subsection{Simulation Setup}
\begin{figure}[!t]
\centering
{\includegraphics[width=7cm]{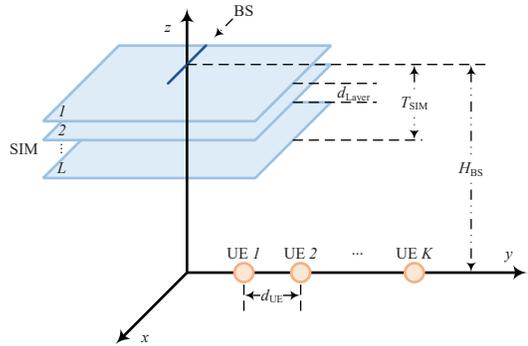}}
\caption{Simulation setup of the considered SIM-assisted multiuser MISO system.}\vspace{-0.6cm}
\label{fig_2}
\end{figure}

In this section, we provide numerical results to validate the effectiveness of the wave-based beamforming design by applying the proposed algorithms. As shown in Fig. \ref{fig_2}, we consider an SIM-assisted multiuser MISO downlink system. The BS is a uniform linear array with $M$ antennas that are deployed parallel to the $x$-axis. Furthermore, an SIM stacking multiple metasurfaces is integrated into the BS to perform the transmit beamforming in the EM wave domain. Each metasurface is a uniform planar array whose meta-atoms are deployed parallel to the $x$-$y$ plane. The center of each antenna/meta-atom of the BS and metasurfaces are located in the $z$-axis. The height of the BS is $H_{\text{BS}} = 10$ m, and the thickness of the SIM is $T_{\text{SIM}} = 5\lambda$. Thus, the spacing between adjacent metasurfaces in a $L$-layer SIM is $d_{\text{Layer}} = T_{\text{SIM}}/L$. Moreover, we assume that all metasurfaces comprise $N_{x}$ and $N_{y}$ meta-atoms along the $x$-axis and $y$-axis, respectively. Thus, the total number of meta-atoms in each layer is $N=N_{x}N_{y}$. For simplicity, we consider a square metasurface structure with $N_{x} = N_{y}$. Furthermore, we assume half-wavelength spacing between adjacent antennas/meta-atoms at the BS and metasurfaces. The size of each meta-atom is $d_{x} = d_{y} = \lambda /2$. We assume that $K$ single-antenna users are evenly distributed on the $y$-axis with spacing $d_{\text{UE}} = 10$ m, as shown in Fig. \ref{fig_2}. Additionally, each antenna at the BS is assumed to have an antenna gain of $5$ dBi, while each user is equipped with a single $0$ dBi antenna \cite{TWC_2022_Papazafeiropoulos_Intelligent}. The distance-dependent path loss is modeled as $\beta_{k}=C_{0}d_{k}^{-\alpha}$, where $C_{0}$ is the free space path loss at the reference distance of $1$ m, and $\alpha$ is the path loss exponent \cite{TCOM_20202_An_Low}. The carrier frequency is $28$ GHz. Also, we set $C_{0} = -60$ dB and $\alpha = 3.5$. The noise power is $\sigma _{k}^{2} = - 104$ dBm for $\forall k \in \mathcal{K}$.

\begin{figure*}[!t]
\centering
\subfloat[]{\includegraphics[width=6cm]{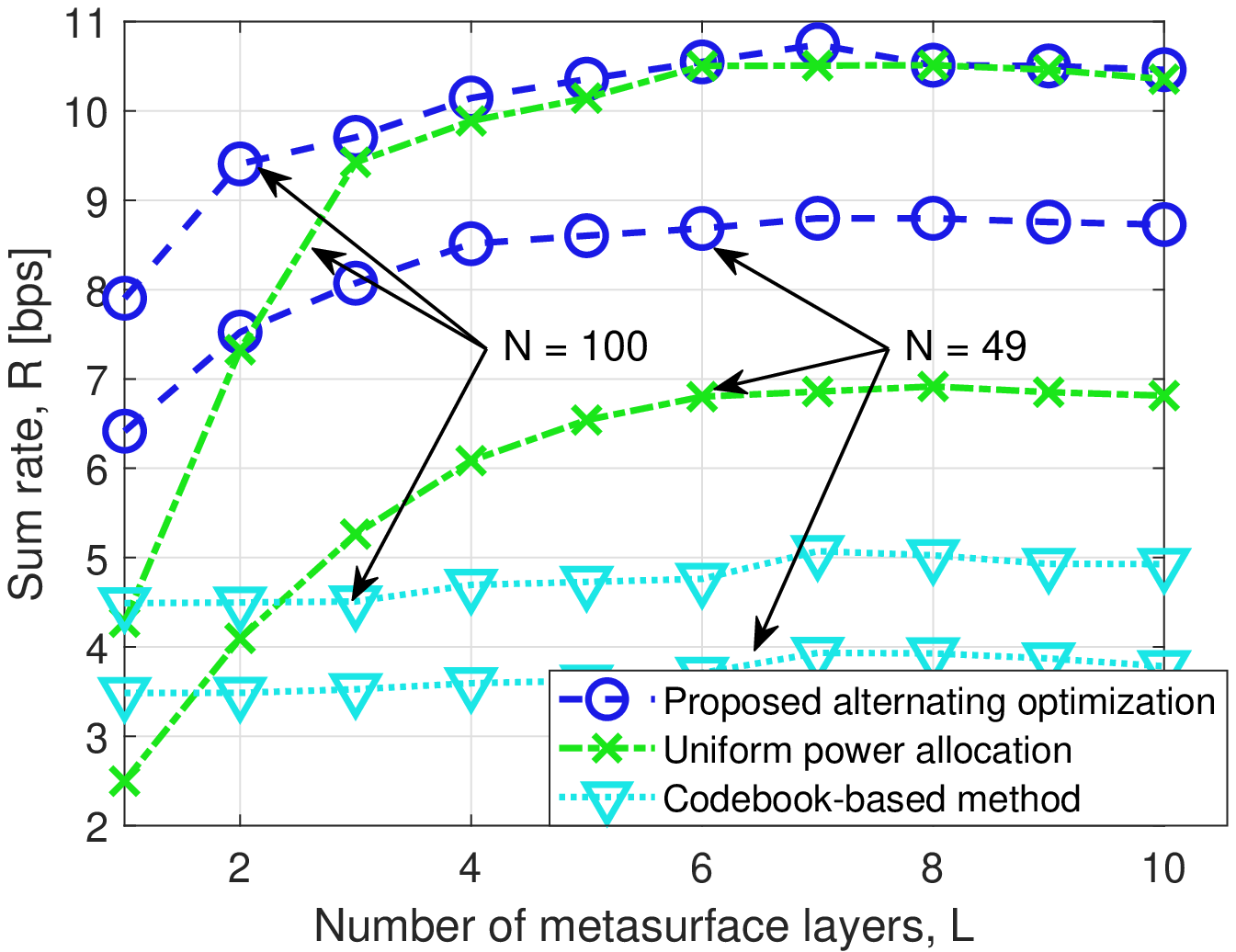}%
\label{fig3_1}}
\hfil
\subfloat[]{\includegraphics[width=6cm]{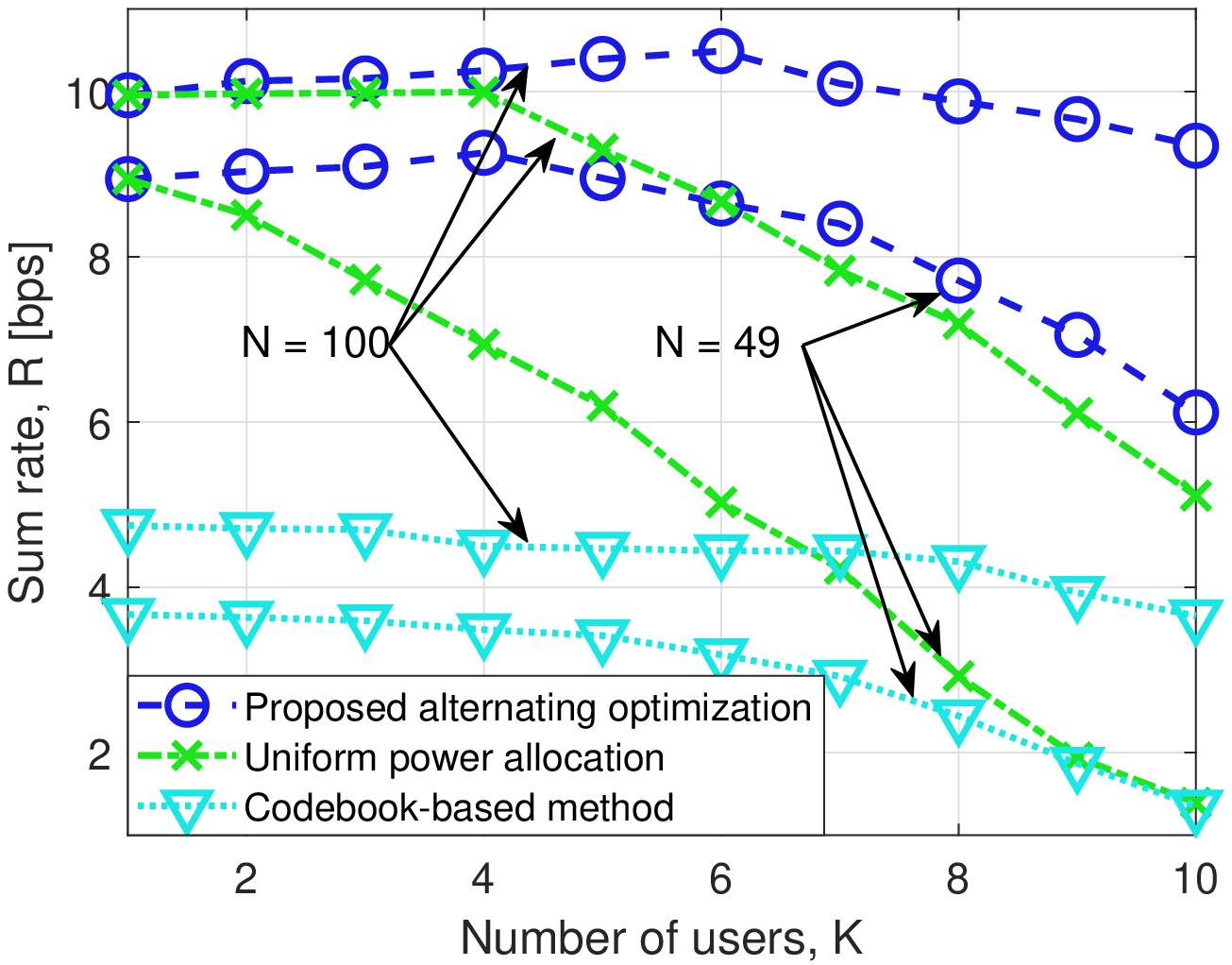}%
\label{fig3_2}}
\caption{(a) Sum rate $R$ versus the number of metasurface layers $L$ ($M = K = 4$, $P_{T} = 10$ dBm); (b) Sum rate $R$ versus the number of users $K$ ($L = 7$, $P_{T} = 10$ dBm);}\vspace{-0.6cm}
\end{figure*}
\begin{figure*}[!t]
\centering
\subfloat[]{\includegraphics[width=6cm]{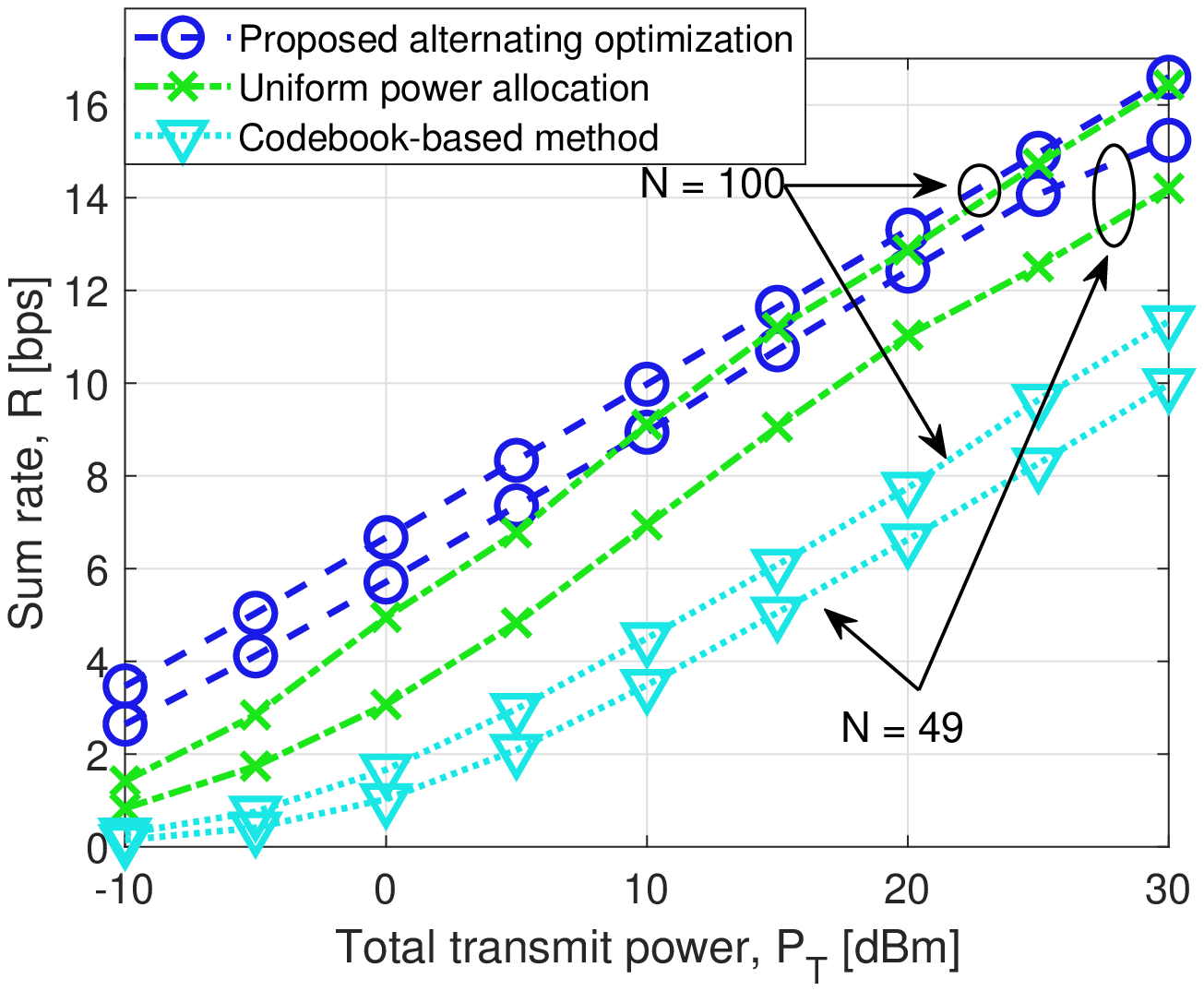}%
\label{fig4_1}}
\hfil
\subfloat[]{\includegraphics[width=6cm]{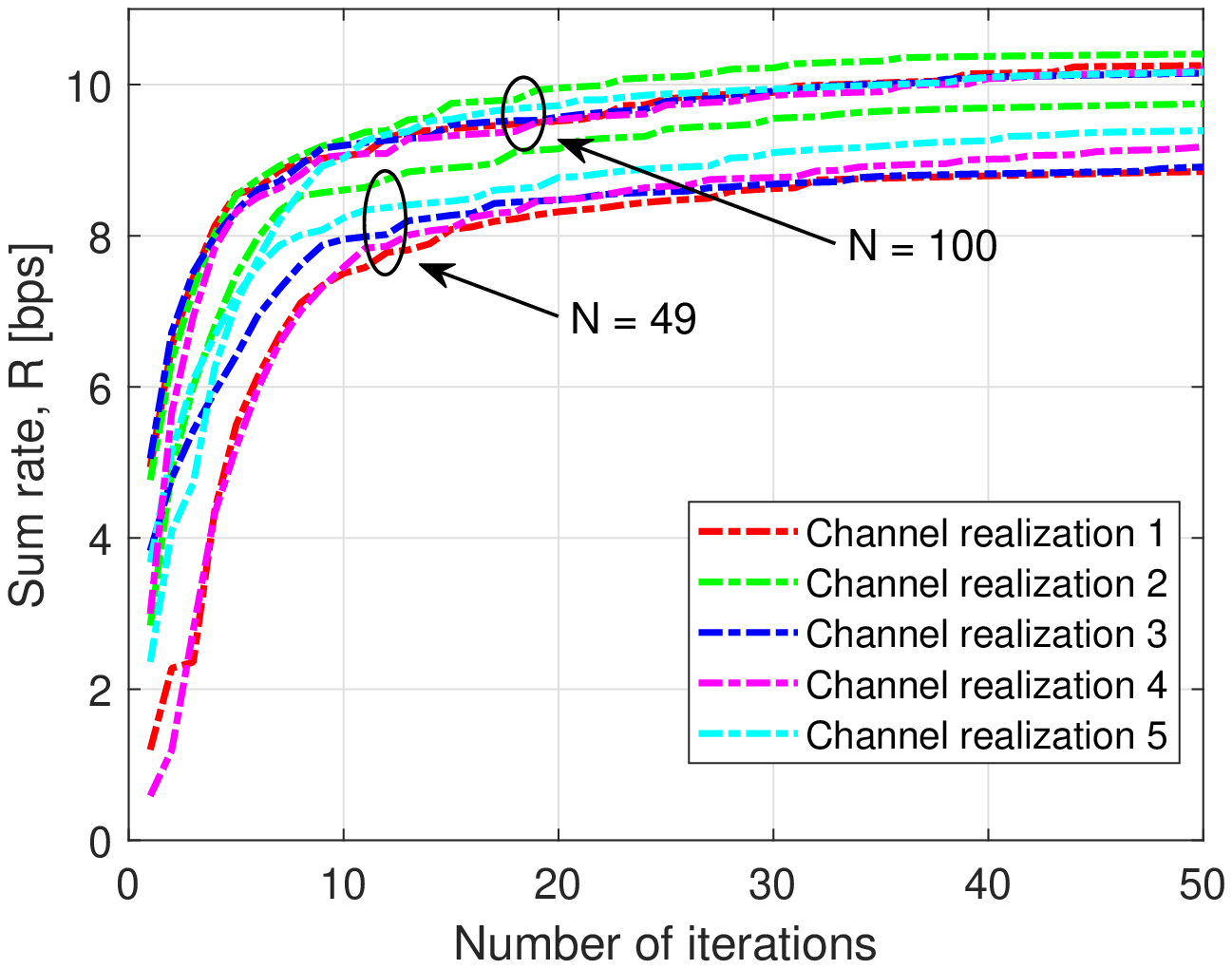}%
\label{fig4_2}}%
\caption{(a) Sum rate $R$ versus the transmit power $P_{T}$ ($L = 7$, $M = K = 4$); (b) Convergence analysis of the AO algorithm.}\vspace{-0.6cm}
\end{figure*}

We evaluate the performance of the proposed algorithm and compare it with two benchmark schemes. 1) Uniform power allocation: This corresponds to solving problem $\left ( P3 \right )$ assuming a uniform power allocation strategy; 2) Codebook-based method: This corresponds to randomly generating a set of phase shift vectors and for each of them applying the iterative water-filling power allocation strategy \cite{TGCN_2022_An_Joint}. The phase shift vector resulting in the maximum sum rate is selected to configure the SIM. The codebook size is $10LN$. The threshold for stopping the AO algorithm is set to $10^{-6}$, and the maximum number of iterations for the AO algorithm and the inner iteration of the iterative water-filling and backtracking line search is set to $100$. The simulation results are obtained by averaging over $100$ independent channel realizations.

\subsection{Sum Rate versus the Number of Metasurface Layers $L$}
In Fig. \ref{fig3_1}, we show the sum rate $R$ versus the number of metasurface layers $L$, by considering the setup $M = K = 4$ and $P_{T} = 10$ dBm. In particular, we analyze two case studies with a different number of meta-atoms in each layer: $N = 49$ and $N = 100$, while increasing the number of metasurface layers $L$ from $ 1$ to $ 10$. We see that the sum rate resulting from the proposed power allocation strategy and wave-based beamforming scheme increases with the number of metasurface layers, benefiting from the capability of an SIM to suppress the multiuser interference in the EM wave domain. Nonetheless, the sum rate gradually converges as $L$ increases, and reaches the maximum at approximately $L=7$, which results in a 30\% improvement compared with a single-layer SIM when $N = 49$. Additionally, the proposed AO algorithm outperforms the codebook-based scheme and the uniform power allocation strategy in all considered setups. Specifically, the uniform power allocation strategy has a rate loss of about $2$ bps compared to the proposed iterative water-filling algorithm when $L \geq 6$. Nonetheless, as the number of meta-atoms $N$ in each layer increases from $49$ to $100$, the sum rate obtained by using the uniform power allocation strategy improves significantly. Notably, the uniform power allocation strategy almost approaches the iterative water-filling power allocation scheme, which means that the multiuser interference is effectively reduced by an SIM as the number of meta-atoms increases. Finally, the codebook-based scheme with random phase shifts does not provide any performance gain as the number of metasurface layers increases. Also, it provides half of the rate compared with the proposed AO algorithm.

\subsection{Sum Rate versus the Number of Users $K$}
In Fig. \ref{fig3_2}, we show the achievable sum rate versus the number of users $K$, by considering the setup $L = 7$. All the other parameters are the same as those in Fig. \ref{fig3_1}. From Fig. \ref{fig3_2}, we observe that the sum rate obtained with the optimized power allocation strategy and wave-based beamforming increases as $K$ increases, it then decreases as $K$ exceeds a certain value. This is due to the fact that a finite-size SIM is capable of hardly suppressing the multiuser interference among a large number of users. In the considered setup with $L = 7$ and $N = 49$, the maximum number of users that an SIM can effectively serve by satisfactorily suppressing the multiuser interference is $K = 4$. The number of users can be increased to $K = 6$ if an SIM with a larger size, e.g., with $N = 100$ meta-atoms in each layer, is utilized. Additionally, the uniform power allocation and the codebook-based schemes hardly suppress the multiuser interference, thus resulting in a noticeable drop of the sum rate as the number of users increases. If the number of users $K$ is large, the iterative water-filling algorithm has a more pronounced performance gain (e.g., 300\% rate improvement when $K = 10$ and $N = 49$) compared with the uniform power allocation strategy, which implies that utilizing an appropriate power allocation algorithm is a suitable approach to further decrease the multiuser interference among a large number of users.

\subsection{Sum Rate versus the Transmit Power $P_T$}
In Fig. \ref{fig4_1}, we illustrate the sum rate versus the transmit power $P_{T}$ by considering a number of users $K = 4$. As expected, the sum rate of all the considered schemes increases as the transmit power $P_{T}$ increases. Again, the proposed AO algorithm outperforms the considered benchmark schemes in all considered setups. Specifically, the performance gain of the proposed algorithm over the codebook-based scheme is more pronounced as the transmit power increases, attaining more than $4.5$ bps rate improvement for moderate and high values of the transmit power. For low values of the transmit power, in addition, the uniform power allocation strategy suffers from some performance loss as compared with the proposed iterative water-filling algorithm. The gap, however, shrinks as the transmit power increases. This is consistent with the typical behavior of the water-filling algorithm in the high power regime, where the impact of power allocation becomes negligible.
\subsection{Convergence Analysis of the Proposed Algorithm}
In Fig. \ref{fig4_2}, we show the convergence of the proposed AO algorithm by considering $N = 49$ and $N =100$, while keeping the other parameters the same as in Fig. \ref{fig3_1}. In each case, we consider $5$ independent channel realizations. Notably, the AO algorithm converges fast in all cases, achieving its maximum value after about $30$ iterations. From Fig. \ref{fig4_2}, we observe that a larger number of iterations is required for reaching convergence as the number of meta-atoms in each layer increases. This is due to the larger number of variables to be optimized and to the larger number of partial derivatives to be computed at each iteration step. Our simulation study reveals, in addition, that the proposed wave-based beamforming scheme with optimized SIM phase shifts can be implemented within a fraction of a nanosecond, which dramatically reduces the processing delay as compared to digital beamforming schemes, which usually require tens of microseconds to be computed.
\section{Conclusions}\label{sec7}
In this paper, we have proposed an SIM-enabled wave-based beamforming design for multiuser MISO downlink systems, which substantially reduces the precoding delay and hardware cost in comparison to its digital counterpart. Specifically, a joint transmit power allocation and phase shift optimization problem has been formulated to maximize the sum rate. Furthermore, we have proposed an effective AO algorithm to decompose the original non-convex joint optimization problem into two subproblems. The power allocation has been tackled by applying a modified iterative water-filling algorithm, while the phase shifts of the SIM have been optimized by customizing the gradient ascent algorithm. Extensive simulation results have demonstrated that the proposed wave-based beamforming design achieves significant performance gains compared to the currently available state-of-the-art benchmarks. Most notably, the wave-based beamforming significantly decreases the precoding delay. In a nutshell, the proposed SIM-enabled transceiver design constitutes a new paradigm to perform advanced signal processing operations in the wave domain.

\bibliography{ref}

\begin{thebibliography}{10}
\providecommand{\url}[1]{#1}
\csname url@samestyle\endcsname
\providecommand{\newblock}{\relax}
\providecommand{\bibinfo}[2]{#2}
\providecommand{\BIBentrySTDinterwordspacing}{\spaceskip=0pt\relax}
\providecommand{\BIBentryALTinterwordstretchfactor}{4}
\providecommand{\BIBentryALTinterwordspacing}{\spaceskip=\fontdimen2\font plus
\BIBentryALTinterwordstretchfactor\fontdimen3\font minus
  \fontdimen4\font\relax}
\providecommand{\BIBforeignlanguage}[2]{{%
\expandafter\ifx\csname l@#1\endcsname\relax
\typeout{** WARNING: IEEEtran.bst: No hyphenation pattern has been}%
\typeout{** loaded for the language `#1'. Using the pattern for}%
\typeout{** the default language instead.}%
\else
\language=\csname l@#1\endcsname
\fi
#2}}
\providecommand{\BIBdecl}{\relax}
\BIBdecl

\bibitem{TCOM_20202_An_Low}
J.~An, C.~Xu, L.~Gan, and L.~Hanzo, ``Low-complexity channel estimation and
  passive beamforming for {RIS}-assisted {MIMO} systems relying on discrete
  phase shifts,'' \emph{IEEE Trans. Commun.}, vol.~70, no.~2, pp. 1245--1260,
  Feb. 2022.

\bibitem{JSAC_2020_Renzo_Smart}
M.~Di~Renzo, A.~Zappone, M.~Debbah, M.-S. Alouini, C.~Yuen, J.~de~Rosny, and
  S.~Tretyakov, ``Smart radio environments empowered by reconfigurable
  intelligent surfaces: How it works, state of research, and the road ahead,''
  \emph{IEEE J. Sel. Areas Commun.}, vol.~38, no.~11, pp. 2450--2525, Nov.
  2020.

\bibitem{TWC_2019_Huang_Reconfigurable}
C.~Huang, A.~Zappone, G.~C. Alexandropoulos, M.~Debbah, and C.~Yuen,
  ``Reconfigurable intelligent surfaces for energy efficiency in wireless
  communication,'' \emph{IEEE Trans. Wireless Commun.}, vol.~18, no.~8, pp.
  4157--4170, Aug. 2019.

\bibitem{Proc_2022_Alexandropoulos_Pervasive}
G.~C. Alexandropoulos, K.~Stylianopoulos, C.~Huang, C.~Yuen, M.~Bennis, and
  M.~Debbah, ``Pervasive machine learning for smart radio environments enabled
  by reconfigurable intelligent surfaces,'' \emph{Proc. IEEE}, vol. 110, no.~9,
  pp. 1494--1525, Sept. 2022.

\bibitem{TC_2021_Wu_Intelligent}
Q.~Wu, S.~Zhang, B.~Zheng, C.~You, and R.~Zhang, ``Intelligent reflecting
  surface-aided wireless communications: A tutorial,'' \emph{IEEE Trans.
  Commun.}, vol.~69, no.~5, pp. 3313--3351, May 2021.

\bibitem{WCL_2022_An_Scalable}
J.~An, Q.~Wu, and C.~Yuen, ``Scalable channel estimation and reflection
  optimization for reconfigurable intelligent surface-enhanced {OFDM}
  systems,'' \emph{IEEE Wireless Commun. Lett.}, vol.~11, no.~4, pp. 796--800,
  Apr. 2022.

\bibitem{WC_2020_Huang_Holographic}
C.~Huang, S.~Hu, G.~C. Alexandropoulos, A.~Zappone, C.~Yuen, R.~Zhang,
  M.~Di~Renzo, and M.~Debbah, ``Holographic {MIMO} surfaces for {6G} wireless
  networks: Opportunities, challenges, and trends,'' \emph{IEEE Wireless
  Commun. Mag.}, vol.~27, no.~5, pp. 118--125, Oct. 2020.

\bibitem{TCOM_2020_Wu_Beamforming}
Q.~Wu and R.~Zhang, ``Beamforming optimization for wireless network aided by
  intelligent reflecting surface with discrete phase shifts,'' \emph{IEEE
  Trans. Commun.}, vol.~68, no.~3, pp. 1838--1851, Mar. 2020.

\bibitem{TWC_2020_Guo_Weighted}
H.~Guo, Y.-C. Liang, J.~Chen, and E.~G. Larsson, ``Weighted sum-rate
  maximization for reconfigurable intelligent surface aided wireless
  networks,'' \emph{IEEE Trans. Wireless Commun.}, vol.~19, no.~5, pp.
  3064--3076, May 2020.

\bibitem{TWC_2022_Papazafeiropoulos_Intelligent}
A.~Papazafeiropoulos, C.~Pan, P.~Kourtessis, S.~Chatzinotas, and J.~M. Senior,
  ``Intelligent reflecting surface-assisted {MU-MISO} systems with imperfect
  hardware: Channel estimation and beamforming design,'' \emph{IEEE Trans.
  Wireless Commun.}, vol.~21, no.~3, pp. 2077--2092, Mar. 2022.

\bibitem{TGCN_2022_An_Joint}
J.~An, C.~Xu, L.~Wang, Y.~Liu, L.~Gan, and L.~Hanzo, ``Joint training of the
  superimposed direct and reflected links in reconfigurable intelligent surface
  assisted multiuser communications,'' \emph{IEEE Trans. Green Commun. Netw.},
  vol.~6, no.~2, pp. 739--754, Jun. 2022.

\bibitem{JSTSP_2022_Wei_Multi}
L.~Wei, C.~Huang, G.~C. Alexandropoulos, W.~E.~I. Sha, Z.~Zhang, M.~Debbah, and
  C.~Yuen, ``Multi-user holographic {MIMO} surfaces: Channel modeling and
  spectral efficiency analysis,'' \emph{IEEE J. Sel. Topics Signal Process.},
  vol.~16, no.~5, pp. 1112--1124, Aug. 2022.

\bibitem{JWCN_2019_Renzo_Smart}
M.~Di~Renzo, M.~Debbah, D.-T. Phan-Huy, A.~Zappone, M.-S. Alouini, C.~Yuen,
  V.~Sciancalepore, G.~C. Alexandropoulos, J.~Hoydis, H.~Gacanin \emph{et~al.},
  ``Smart radio environments empowered by reconfigurable {AI} meta-surfaces: An
  idea whose time has come,'' \emph{EURASIP J. Wireless Commun. Netw.}, vol.
  2019, no.~1, pp. 1--20, May 2019.

\bibitem{WC_2022_An_Codebook}
J.~An, C.~Xu, Q.~Wu, D.~W.~K. Ng, M.~Di~Renzo, C.~Yuen, and L.~Hanzo,
  ``Codebook-based solutions for reconfigurable intelligent surfaces and their
  open challenges,'' \emph{IEEE Wireless Commun.}, pp. 1--8, Early Access,
  2022.

\bibitem{Science_2018_Lin_All}
X.~Lin, Y.~Rivenson, N.~T. Yardimci, M.~Veli, Y.~Luo, M.~Jarrahi, and A.~Ozcan,
  ``All-optical machine learning using diffractive deep neural networks,''
  \emph{Sci.}, vol. 361, no. 6406, pp. 1004--1008, Jul. 2018.

\bibitem{NE_2022_Liu_A}
C.~Liu, Q.~Ma, Z.~J. Luo, Q.~R. Hong, Q.~Xiao, H.~C. Zhang, L.~Miao, W.~M. Yu,
  Q.~Cheng, L.~Li \emph{et~al.}, ``A programmable diffractive deep neural
  network based on a digital-coding metasurface array,'' \emph{Nat. Electro.},
  vol.~5, no.~2, pp. 113--122, Feb. 2022.

\bibitem{TCOM_2022_Liu_Compact}
K.~Liu, Z.~Zhang, L.~Dai, and L.~Hanzo, ``Compact user-specific reconfigurable
  intelligent surfaces for uplink transmission,'' \emph{IEEE Trans. Commun.},
  vol.~70, no.~1, pp. 680--692, Jan. 2022.

\bibitem{WCL_2021_Bjornson_Rayleigh}
E.~Björnson and L.~Sanguinetti, ``Rayleigh fading modeling and channel
  hardening for reconfigurable intelligent surfaces,'' \emph{IEEE Wireless
  Commun. Lett.}, vol.~10, no.~4, pp. 830--834, Apr. 2021.

\bibitem{CM_2004_Sanayei_Antenna}
S.~Sanayei and A.~Nosratinia, ``Antenna selection in {MIMO} systems,''
  \emph{IEEE Commun. Mag.}, vol.~42, no.~10, pp. 68--73, Oct. 2004.

\bibitem{TCOM_2016_Lee_Channel}
J.~Lee, G.-T. Gil, and Y.~H. Lee, ``Channel estimation via orthogonal matching
  pursuit for hybrid {MIMO} systems in millimeter wave communications,''
  \emph{IEEE Trans. Commun.}, vol.~64, no.~6, pp. 2370--2386, June 2016.

\bibitem{TIT_2005_Jindal_Sum}
N.~Jindal, W.~Rhee, S.~Vishwanath, S.~Jafar, and A.~Goldsmith, ``Sum power
  iterative water-filling for multi-antenna {Gaussian} broadcast channels,''
  \emph{IEEE Trans. Inf. Theory}, vol.~51, no.~4, pp. 1570--1580, Apr. 2005.

\end{thebibliography}
\bibliographystyle{IEEEtran}
\end{document}